\documentclass[aps,prd,reprint,preprintnumbers,superscriptaddress,showpacs,twocolumn]{revtex4-1}
\usepackage{latexsym,graphicx,amssymb,amsmath,mathrsfs}
\usepackage{setspace,bm}
\usepackage[breaklinks, colorlinks=true, pdfstartview=FitV, linkcolor=red, citecolor=blue, urlcolor=blue]{hyperref}
\usepackage[usenames]{color}
\usepackage{latexsym}
\usepackage{epstopdf}
\usepackage{mathtools}
\usepackage{url}
\usepackage{comment}
\usepackage{braket}
\usepackage{CJKutf8}
\usepackage{physics}
\usepackage{slashed}
\usepackage{bbold}

\begin{document}

\title{Mixed-state quantum geometry of dense two-color QCD matter}

\author{Kouji Kashiwa}
\email[]{kashiwa@fit.ac.jp}
\affiliation{Department of Computer Science and Engineering, Faculty of Information Engineering, Fukuoka Institute of Technology, Fukuoka 811-0295, Japan}

\author{Riki Oshima}
\affiliation{Department of Physics, Saga University,
             Saga 840-8502, Japan}
             
\author{Hiroaki Kouno}
\affiliation{Department of Physics, Saga University,
             Saga 840-8502, Japan}

\begin{abstract}
We investigate an information-geometric approach to dense two-color QCD using thermal states defined as normalized Gibbs states in the two-flavor Nambu--Jona-Lasinio model at the mean-field level.
The Bures metric and quantum Fisher information are constructed from thermal states associated with effective Bogoliubov--de Gennes Hamiltonians reproducing the mean-field quasiparticle spectrum.
On the thermodynamic parameter manifold $(T,\mu)$, where $T$ is temperature and $\mu$ is quark chemical potential, we analyze information-geometric quantities including the trace of the QFI matrix and the metric-volume density.
The numerical results show pronounced enhancements of both quantities near the color-superconducting phase boundary, whereas only weaker signatures are observed across the chiral crossover.
By contrast, neither observable exhibits a distinct feature associated with the BEC--BCS crossover.
\end{abstract}

\maketitle

\section{Introduction}
\label{sec:introduction}
Understanding the properties of quantum chromodynamics (QCD) at finite temperature ($T$) and quark chemical potential ($\mu$) is important for understanding our universe.
However, at finite $\mu$, QCD suffers from the sign problem, and thus the first-principles lattice QCD simulation becomes difficult; for example, see Refs.\,\,\cite{deForcrand:2010ys,Nagata:2021bru,Nagata:2021ugx}.
In contrast, two-color QCD, whose gauge group is $SU(2)$, provides a useful QCD-like theory because the pseudo-reality of the fundamental representation removes the sign problem~\cite{Kogut:2000ek}: for details, see Ref.\,\cite{Fukushima:2007bj} as an example.
Therefore, it provides a controlled theoretical laboratory in which we can investigate properties of dense QCD matter, including chiral symmetry breaking, diquark condensation, and crossover from a Bose--Einstein condensate (BEC) to a Bardeen--Cooper--Schrieffer (BCS) state: for example, see Ref.\,\cite{Hands:2007by}.

A distinctive feature of two-color QCD is that baryons are bosonic diquarks.
At sufficiently large $\mu$, the ground state can develop a scalar diquark condensate, $\langle q^T C i\gamma_5\tau_2 t_2 q \rangle$, where $q$ denotes the quark field, $C$ is the charge-conjugation matrix, and the Pauli matrix $\tau_2$ ($t_2$) acts in the flavor (color) space.
The diquark condensate and the chiral condensate $\sigma = \langle \bar q q \rangle$ are natural competing order parameters on the QCD phase diagram at finite $T$ and $\mu$; this type of diquark condensate can also appear in three-color QCD~\cite{Alford:1997zt,Rapp:1997zu}.
In contrast, at sufficiently high density, three-color QCD is expected to realize a different type of diquark condensation, such as the color-flavor-locking (CFL) \,\cite{Alford:1998mk,Schafer:1998ef,Evans:1999at}; see Refs.\,\cite{Alford:2007xm,Alford:2001dt} as a review.
Although the diquark pairing structure in two-color QCD somewhat differs from that of three-color QCD, the diquark condensate plays an analogous role and provides a valuable framework for exploring the QCD phase diagram.

Information geometry provides an alternative description of thermodynamic states in terms of the distinguishability of thermal states rather than order parameters; for example, see Ref.\,\cite{Gu:2008dyh}.
In this study, we focus on the Bures metric~\cite{bures1969extension} and the quantum Fisher information (QFI)~\cite{helstrom1969quantum,Braunstein:1994zz}, which quantify the distinguishability of neighboring thermal states.
Both quantities are closely related to the Uhlmann formulation of mixed-state distinguishability~\cite{Uhlmann:1975kt}.
These quantities are constructed directly from the thermal state and its parameter derivatives.
Conventional susceptibilities are usually defined from derivatives of a particular thermodynamic quantity or order parameter; see Appendix~\ref{susceptibility} as an example.
By contrast, the QFI measures the distinguishability of neighboring mixed states through the thermal state itself.
Therefore, it reflects changes in the thermal state rather than fluctuations of a single order parameter.

Geometric approaches to QCD matter have also been extensively explored from a thermodynamic perspective.
Within the framework of Ruppeiner geometry~\cite{Ruppeiner:1995zz}, a metric is constructed on the space of equilibrium thermodynamic states from thermodynamic fluctuations, and the associated scalar curvature has been proposed as a useful probe of phase transitions and critical phenomena~\cite{Castorina:2018ayy,Zhang:2019neb,Oshima:2026bub}.
In particular, the thermodynamic curvature has been applied to the QCD phase diagram, where both the sign and magnitude of the curvature have been used to characterize crossover behavior, critical fluctuations, and the vicinity of the critical endpoint.
The present work adopts a different geometric approach from that of Ruppeiner geometry.
Rather than constructing geometry from thermodynamic potentials, we employ a mixed-state quantum geometric framework based on the thermal state itself.

The main questions addressed in this study are the following:
\begin{enumerate} 
    \item Can information-geometric quantities derived from the thermal state provide useful diagnostics of dense two-color QCD matter?
    \item How are these quantities related to diquark condensation, the chiral crossover, and the BEC--BCS crossover?
\end{enumerate}
Note that the present formulation should be understood as a mean-field quasiparticle information geometry of dense two-color QCD matter, because we employ the mean-field approximation in this study.
The two-flavor and two-color Nambu--Jona-Lasinio (NJL) model is employed as an effective laboratory for investigating dense two-color QCD matter as a first step.
The present study therefore does not aim to provide the full nonperturbative information geometry of QCD.
Instead, we study the information geometry of normalized Gibbs states associated with effective BdG Hamiltonians whose parameters are determined from the mean-field NJL thermodynamic potential.

This paper is organized as follows.
In the next section, we explain the formulation of the NJL model and information geometric quantities.
Section \ref{sec:numerical_results} shows the numerical results, and Sec.\,\ref{sec:discussions} and \ref{sec:summary} are devoted to discussion and summary, respectively.

\section{Formulation}
\label{sec:formulation}
In this section, we first explain the two-flavor and two-color NJL model, which includes the quark-quark interaction.
Second, we show the reduced Hamiltonian based on the Bogoliubov--de Gennes (BdG) representation and the thermal state.
Finally, information geometric quantities such as the QFI and the metric-volume density are explained.

\subsection{NJL model with diquark gap}
In this study, we consider a two-flavor and two-color Nambu--Jona-Lasinio (NJL) model with the scalar-type quark-quark interaction;
\begin{align}
    {\mathcal L}
    &=\bar{q} \left(i\gamma_\mu\partial^\mu - m_0 + \mu \gamma^0 \right) q
      +G_\mathrm{S}\left[(\bar q q)^2 + (\bar q i \gamma_5 \bm{\tau} q)^2\right]
    \nonumber\\
    &\quad
      +G_\mathrm{D}
      \left(\bar{q} i \gamma_5 \tau_2 t_2 C \bar{q}^T \right)
      \left(q^T C i \gamma_5 \tau_2 t_2 q \right),
    \label{eq:NJL}
\end{align}
where $\gamma$ denote the gamma matrices, $G_\mathrm{S}$ and $G_\mathrm{D}$ are the coupling constants for each interaction channel, $m_0$ denotes the current quark mass; see Ref.\,\cite{Huang:2004ik} for details of the two-flavor color-superconductivity and its effective model.
The chiral condensate $\sigma$ and the diquark gap $\Delta$ that corresponds to the diquark condensate $\langle q^T C i\gamma_5\tau_2t_2 q\rangle$ including the coupling constant $G_\mathrm{D}$, are defined as
\begin{align}
    \sigma &= \langle \bar q q\rangle,\\
    \Delta &= -2G_\mathrm{D} \langle q^T C i\gamma_5\tau_2t_2 q\rangle.
\end{align}
By introducing the Nambu--Gor'kov spinor $\Psi$ as
\begin{align}
    \Psi(p)=
    \begin{pmatrix}
        q(p)\\ q^C(-p)
    \end{pmatrix},
    \qquad
    q^C=C\bar q^T,
\end{align}
the Lagrangian density under the mean-field approximation ${\mathcal L}_\mathrm{MF}$ becomes
\begin{align}
    {\mathcal L}_\mathrm{MF}
    = \frac{1}{2} \bar{\Psi}(p) S^{-1}_{\mathrm{NG}}(p) \Psi(p)
    - \frac{(M-m_0)^2}{4G_\mathrm{S}} - \frac{|\Delta|^2}{4G_\mathrm{D}}.
\end{align}
where the constituent quark mass $M$ is defined as
\begin{align}
    M &= m_0 - 2 G_\mathrm{S} \sigma,
    \label{eq:M_sigma_Bures}
\end{align}
and the inverse Nambu--Gorkov propagator $S^{-1}_\mathrm{NG}$ has a block structure such as
\begin{align}
    S^{-1}_\mathrm{NG}(p)=
    \begin{pmatrix}
        \slashed p + \mu \gamma^0 - M & i \gamma_5 \tau_2 t_2 \Delta \\
        i \gamma_5 \tau_2 t_2 \Delta^* & \slashed p - \mu\gamma^0 - M
    \end{pmatrix},
    \label{eq:SNG_inverse}
\end{align}
here $\slashed p = \gamma_\mu p^\mu$ with the four dimensional momentum $p^\mu$.

The mean-field thermodynamic potential $\Omega_\mathrm{MF}$ is obtained as
\begin{align}
    \Omega_\mathrm{MF}(T,\mu;M,\Delta)
    &= \frac{(M-m_0)^2}{4G_\mathrm{S}} + \frac{|\Delta|^2}{4G_\mathrm{D}}
    \nonumber\\
    &- \frac{T}{2V_3} \ln \det S^{-1}_\mathrm{NG},
    \label{eq:Omega_NJL}
\end{align}
where $V_3$ denotes the spatial volume of the system.
Note that the NJL model is not renormalizable and thus we introduce the three-dimensional momentum cutoff $\Lambda$ for the momentum integration appearing in Eq.\,(\ref{eq:Omega_NJL}).
After taking the Matsubara sum and internal traces in the imaginary time formalism, the quasiparticle energies appear as
\begin{align}
    E_p &= \sqrt{\bm{p}^{2} + M^2},\\
    E_\Delta^\pm(p) &= \sqrt{(E_p \pm \mu)^2 + |\Delta|^2}.
    \label{eq:dispertion}
\end{align}
The values of the mean fields, $\sigma$ and $\Delta$, are obtained by solving the gap equations:
\begin{align}
    \frac{\partial\Omega_\mathrm{MF}}{\partial M}=0,
    \qquad
    \frac{\partial\Omega_\mathrm{MF}}{\partial \Delta}=0,
    \label{eq:gap_Bures}
\end{align}
For simplicity, a real diquark gap is assumed here, and thus we impose it below.
The parameters are summarized in Table\,\ref{tab:parameter}; see Refs.\,\cite{Brauner:2009gu,Kashiwa:2012xm} for details of the parameters and the phase diagram.
Note that $G_\mathrm{D}=G_\mathrm{S}$ is imposed due to the Pauli--G\"ursey symmetry; for example, see Ref.\,\cite{Ratti:2004ra}.
\begin{table}[t]
\caption{Parameters in the NJL model.}
\vspace{2mm}
\centering
    \begin{tabular}{ccccc}
    \hline
        $m_0$ [MeV] & $\Lambda$ [MeV] & $G_\mathrm{S}$ [GeV$^{-2}$] &
        $G_\mathrm{D}/G_\mathrm{S}$ & $M_0$ [MeV] \\
    \hline
        5.4 & 657 & 7.23 & 1.0 & 305 \\
    \hline
\end{tabular}
\label{tab:parameter}
\end{table}

\begin{figure*}[t]
    \centering
    \includegraphics[width=0.49\linewidth]{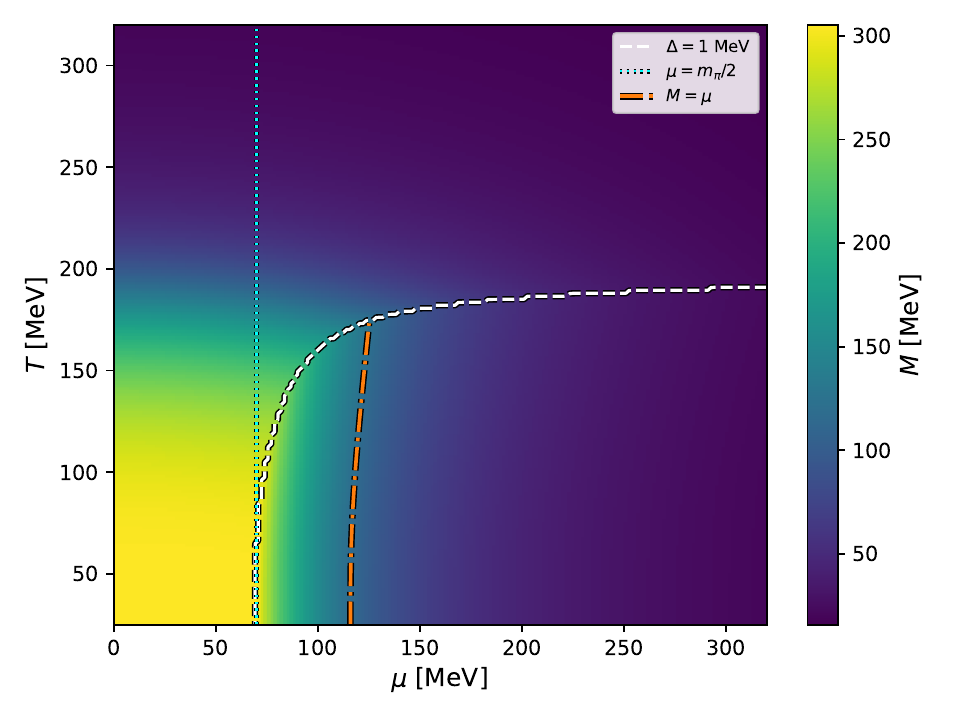}
    \includegraphics[width=0.49\linewidth]{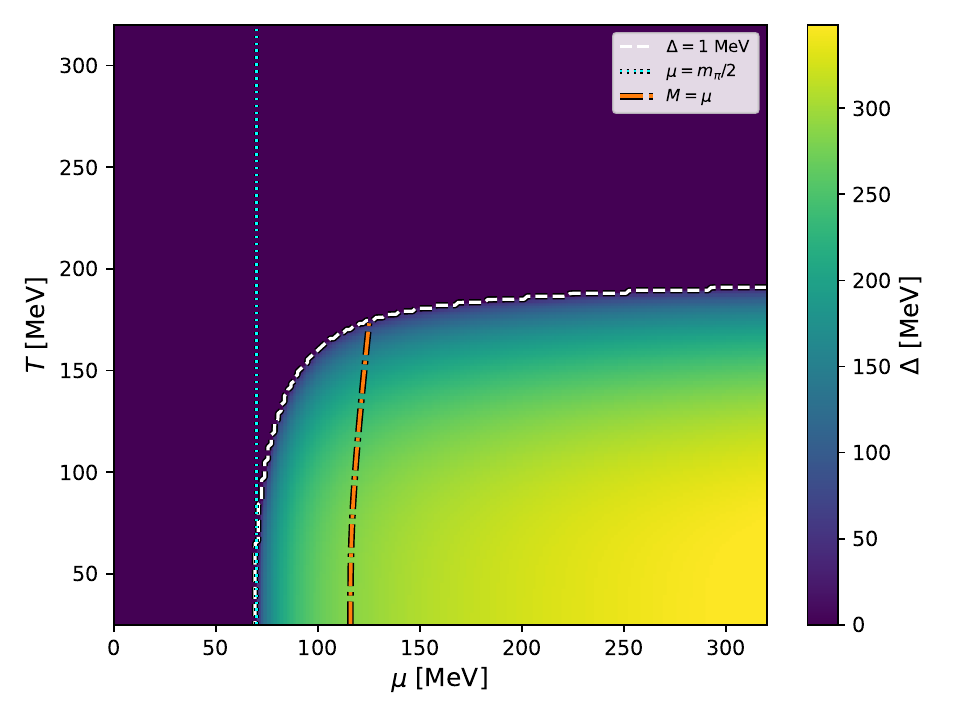}
    \caption{The constituent quark mass and diquark gap in the $T$--$\mu$ plane.
    The dashed line indicates the color-superconducting phase boundary, for which $\Delta = 1~\mathrm{MeV}$ is used as a practical criterion.
    The vertical dotted line shows $\mu=m_\pi/2$, which corresponds to the expected onset of superfluidity in the two-color QCD in the low-$T$ limit.
    The value of $m_\pi$ is set to $140$ MeV.
    Since no true phase transition separates the BEC and BCS regimes, following the conventional criterion~\cite{Sun:2007fc}, we employ condition $\mu = M$ as a practical indicator of the BEC--BCS crossover inside the color-superconducting phase.}
    \label{fig:M_Delta}
\end{figure*}
The left and right panels of Fig.~\ref{fig:M_Delta} show the constituent quark mass $M$ and the diquark gap $\Delta$ in the $T$--$\mu$ plane, respectively.
The constituent mass gradually decreases toward the high-density region, reflecting the partial restoration of chiral symmetry.
At small $\mu$, the chiral crossover occurs around
$T \sim 200 ~ \mathrm{MeV}$, where the constituent quark mass $M$
decreases smoothly with increasing $T$.
In contrast, the diquark condensate develops in the low-temperature and high-density regime, indicating the onset of diquark condensation.

\subsection{Hamiltonian and associated Gibbs state}
The Hamiltonian follows from the equation of motion associated with the inverse Nambu--Gor'kov propagator (\ref{eq:SNG_inverse}).
Changing $p_0$ by $i\partial_t$ and multiplying the equation by the Nambu--Gor'kov extension of $\gamma^0$, the quasiparticle equation can be cast into the Schr\"odinger equation
\begin{align}
    i \frac{\partial \Psi}{\partial t} &= H_\mathrm{NG}\Psi,
\end{align}
with the mean-field Hamiltonian $H_\mathrm{NG}$, which is Hermitian,
\begin{align}
H_\mathrm{NG} &=
    \begin{pmatrix}
        \bm{\alpha} \cdot \bm{p} + \beta_D M - \mu
      & \beta_D i \gamma_5 \tau_2 t_2 \Delta \\
        \beta_D i\gamma_5\tau_2 t_2 \Delta
      & \bm{\alpha} \cdot \bm{p} + \beta_D M + \mu
\end{pmatrix},
\end{align}
where $\alpha_i=\gamma^0\gamma_i$ and $\beta_D=\gamma^0$.   
Introducing the eigenbasis of the Dirac Hamiltonian $H_\mathrm{D}$ as
\begin{align}
    H_\mathrm{D} u_\pm(p)=\pm E_p u_\pm(p),
\end{align}
where $H_\mathrm{D}=\bm{\alpha} \cdot \bm{p} + \beta_D M$, $ H_\mathrm{NG}$ can be projected into the positive- and negative-energy sectors, separately.
After taking into account the color--flavor structure associated with the diquark pairing and diagonalizing the Dirac sector, the quasiparticle spectrum is described by an effective Bogoliubov-de Gennes (BdG) Hamiltonian $H_\pm(p)$ as
\begin{align}
    H_\pm(p)
    &=
    \begin{pmatrix}
        \xi_\pm(p) & \Delta\\
        \Delta     & - \xi_\pm(p)
    \end{pmatrix}
    = \xi_\pm(p) \sigma_z + \Delta \sigma_x,
    \label{eq:Hamiltonian_BdG}
\end{align}
where $ \xi_\pm(p) = E_p\pm \mu$.
This is precisely the Bogoliubov--de Gennes Hamiltonian for a paired fermionic system.
The eigenvalues of $H_\pm(p)$ are
\begin{align}
    \lambda_\pm
    &= \pm \sqrt{\xi_\pm(p)^2 + \Delta^2}
     = \pm E_\Delta^\pm(p),
\end{align}
see Ref.\,\cite{de2018superconductivity} for details of the BdG Hamiltonian.
Therefore, the reduced Hamiltonian reproduces exactly the quasiparticle dispersion relation (\ref{eq:dispertion}), while retaining the pairing structure generated by the diquark gap; for more details, see Appendix~\ref{Justification}.

In this study, we define the associated quasiparticle thermal state as the normalized Gibbs state.
For each momentum mode, the thermal state $\rho_\mathrm{NG}(\bm{p};T,\mu)$ associated with the mean-field Nambu--Gor'kov Hamiltonian is defined as
\begin{align}
    \rho_\mathrm{NG}(\bm{p};T,\mu)
    =\frac{\exp[-H_\mathrm{NG}(\bm{p};T,\mu)/T]}
           {\operatorname{Tr} \exp[-H_\mathrm{NG}(\bm{p};T,\mu)/T]},
    \label{eq:rho}
\end{align}
where $\operatorname{Tr}$ denotes the trace over the Dirac, color, flavor and Nambu--Gor'kov indices.
Equation (\ref{eq:rho}) illustrates the normalized Gibbs-state construction at the level of the mean-field Nambu--Gor'kov Hamiltonian.

From the spectral decomposition of $H_\mathrm{NG}$:
\begin{align}
    H_\mathrm{NG} | n \rangle &= \varepsilon_n | n \rangle,
\end{align}
the quasiparticle thermal state can then be written as
\begin{align}
    \rho_\mathrm{NG} &= \sum_n r_n |n\rangle \langle n|,
\end{align}
with
\begin{align}
    r_n &= \frac{e^{-\varepsilon_n/T}} {\sum\limits_m e^{-\varepsilon_m/T}}.
\end{align}
Since the thermal state is positive definite, we have $r_n>0$ and, therefore, $r_m+r_n>0$.
Although the explicit diagonalization of the Hamiltonian is straightforward, the following expressions are written in terms of its generic spectral decomposition.

In the numerical implementation, we utilize the block-diagonal structure of the mean-field Nambu--Gor'kov Hamiltonian.
After diagonalizing the fixed internal color, flavor, Dirac and Nambu--Gor'kov spaces, the quasiparticle spectrum is represented by the two BdG branches $E_\Delta^\pm(p)$.
For each BdG branch, we use the minimal $2 \times 2$ BdG block (\ref{eq:Hamiltonian_BdG}), whose eigenvalues reproduce $\pm E_\Delta^\pm (p)$.
The remaining internal degeneracies are included through a degeneracy factor in the momentum integral, as explained in the next section. 
The normalized Gibbs state of the reduced BdG matrix should not be identified with the full many-body Gibbs density operator in fermionic Fock space.
Rather, it defines an effective quasiparticle-state manifold whose information geometry is investigated here.
This reduced BdG block reproduces $E_\Delta^\pm(p)$ and retains the essential diquark pairing structure.
A full calculation using the complete Nambu--Gor'kov Hamiltonian would be desirable for checking the quantitative robustness of the information-geometric observables.
It is our future work.

\subsection{Bures metric and quantum Fisher information}
In this section, we explain the Bures metric and the quantum Fisher information~\cite{bures1969extension,Uhlmann:1986fko}.
The QFI plays a central role in quantum estimation theory and quantum information science~\cite{Braunstein:1994zz,liu2020quantum}.
Recently, the quantum geometry of BdG quasiparticle states has attracted attention in superconducting systems~\cite{kaplan2025quantum,simon2026composite,keskiner2026superconductivity}.

The Uhlmann fidelity between two thermal states is
\begin{align}
    \mathcal F(\rho_\mathrm{A},\rho_\mathrm{B})
    &= \left[\operatorname{Tr} \sqrt{ \sqrt{\rho_\mathrm{A}} \rho_\mathrm{B} \sqrt{\rho_\mathrm{A}} } \right]^2,
    \label{eq:Uhlmann_F}
\end{align}
and the Bures distance is then defined as
\begin{align}
    ds_\mathrm{Bures}^2 &= 2\left(1-\sqrt{\mathcal F(\rho,\rho + d\rho)} \right).
    \label{eq:Bures_distance}
\end{align}
For infinitesimally separated states on the manifold, we can write
\begin{align}
    ds_\mathrm{Bures}^2 &= g^{(\mathrm{Bures})}_{ij} dx^i dx^j,
    \label{eq:ds_bures}
\end{align}
where $x^i=(T,\mu)$.
With the spectral basis of $\rho$, $g_{ij}^{(\mathrm{B})}$ becomes
\begin{align}
    g^{(\mathrm{Bures})}_{ij}
    =\frac{1}{2}\sum_{m,n}
      \frac{\bra{m} \partial_i \rho \ket{n}
            \bra{n} \partial_j \rho \ket{m}}{r_m+r_n},
    \label{eq:Bures_metric}
\end{align}
for details, see Ref.\,\cite{Hubner:2002wza}.
Since the thermal state is positive in the present setting, the denominator is nonzero in the numerical calculation.
This fidelity-based approach can be used to characterize critical phenomena and quantum phase transitions~\cite{zanardi2007information,Venuti:2007izd}.

The QFI metric is related to the Bures metric~\cite{Braunstein:1994zz,Paris:2008zgg} as
\begin{align}
    g^{(\mathrm{QFI})}_{ij} &= 4 g^{(\mathrm{Bures})}_{ij}.
    \label{eq:QFI_metric_Bures}
\end{align}
Beyond its original applications in quantum metrology, the QFI has become a useful diagnostic for many-body systems and phase transitions~\cite{zanardi2007information,Venuti:2007izd}.

For visualization in this study, we use the coordinate trace:
\begin{align}
    g^{(\mathrm{QFI})}_{TT} + g^{(\mathrm{QFI})}_{\mu\mu},
    \label{eq:Q_trace}
\end{align}
which provides a simple measure of the overall local distinguishability of neighboring thermal states in the $T$--$\mu$ plane.
Throughout this study, we use (\ref{eq:Q_trace}) because it provides a simple measure of the overall local distinguishability.
In the following, we call the quantity (\ref{eq:Q_trace}) the QFI trace after integrating it over momentum.

The thermal state with the effective BdG Hamiltonian (\ref{eq:Hamiltonian_BdG}) becomes
\begin{align}
    \rho_{\pm}(\bm{p})
    &= \frac{e^{-H_{\pm}/T}}{\operatorname{Tr} \, e^{-H_{\pm}/T}}.
    \label{eq:density_matrix_pm}
\end{align}
For details of the thermal state, see Appendix~\ref{Bloch_representation}.
Since $H_\pm(p)$ acts on a two-dimensional Nambu--Gor'kov space, the thermal state is completely specified by a $2\times2$ matrix.
The QFI metric admits the following expression;
\begin{align}
    g^{(\mathrm{QFI})}_{ij}
    &= \partial_i \bm{r} \cdot \partial_j \bm{r}
     + \frac{ (\bm{r} \cdot \partial_i \bm{r})(\bm{r} \cdot \partial_j \bm{r})}
            { 1 - |\bm{r} |^2 }.
\label{eq:QFI_Bloch}
\end{align}
For its derivation, see Appendix~\ref{QFI_ap}.
Thus, the QFI is controlled by the response of the Bloch vector $\bm{r}$ to the thermodynamic parameters.

In the numerical calculation, we evaluate the QFI metric of each BdG branch using Eq.\,(\ref{eq:QFI_Bloch}) rather than the general eigenstate representation~(\ref{eq:QFI_metric_Bures}).
The two formulations are mathematically equivalent for a two-level thermal state.
The Bloch-vector representation is numerically advantageous because it avoids repeated diagonalization of nearly degenerate thermal states.
The total information metric is then obtained from the momentum integration and branch summation.
The derivatives of the mean fields with respect to $T$ and $\mu$, which appear in the metric, were numerically evaluated using the local polynomial fitting on the $(T,\mu)$ grid to control the numerical uncertainties.
Note that we encounter a numerical difficulty in the evaluation of the metric in the low-$T$ region.

At sufficiently low $T$, the thermal state approaches a nearly pure-state limit $|\bm{r}| \rightarrow 1$.
Consequently, the denominator $1-|\bm{r}|^2 = \sech^2 (E^\Delta_\pm/T)$ in Eq.\,(\ref{eq:QFI_Bloch}) becomes very small and the metric becomes increasingly sensitive to small numerical uncertainties in the derivatives of $\bm{r}$.
This effect is particularly severe for the metric-volume density, since the determinant evaluation can amplify small numerical errors.
Therefore, we restrict the numerical analysis to $T \ge 25$ MeV. 

\subsection{Momentum integration}
\begin{figure*}[t]
    \centering
    \includegraphics[width=0.49\linewidth]{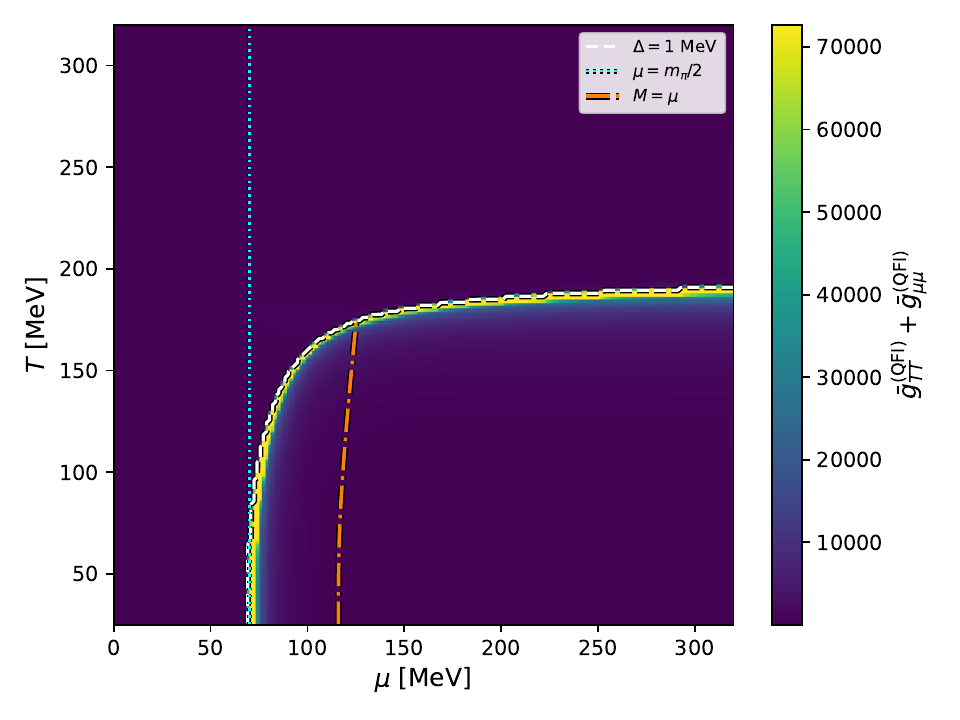}
    \includegraphics[width=0.49\linewidth]{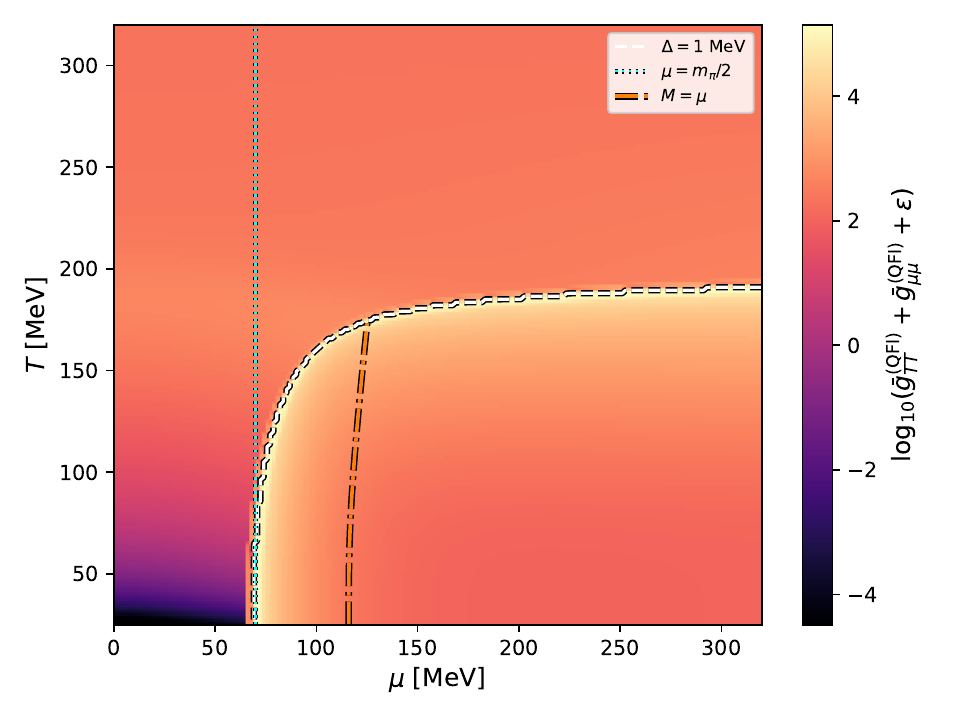}
    \caption{The QFI trace in the $T$--$\mu$ plane.
    The left and right panels show the linear and log scale results, respectively.
    The sufficiently small positive value $\epsilon$ is introduced to avoid zero inside the logarithm.
    The dashed and dotted lines are the same as those shown in Fig.\,\ref{fig:M_Delta}.}
    \label{fig:QFI}
\end{figure*}

Within the mean-field quasiparticle representation, each momentum block is further decomposed into the two quasiparticle branches $E_\Delta^\pm(p)$.
Because the mean-field Hamiltonian is quadratic and block-diagonal in momentum space, different momentum sectors are independent, allowing the thermal state to factorize into momentum-resolved BdG blocks.
The effective quasiparticle state over the momentum-resolved BdG blocks is then constructed as
\begin{align}
        \rho_{\mathrm{Total}}
        &= \bigotimes_{\bm{p} \in \mathcal{P}}\rho_{\bm{p}},
\end{align}
where
\begin{align}
        \rho_{\bm{p}} 
        &= \rho_+(\bm{p}) \otimes \rho_-(\bm{p}),
\end{align}
here ${\cal P}$ denotes a set of independent momentum blocks.
This factorization should be regarded as a property of the mean-field BdG quasiparticle description rather than of the full interacting QCD thermal state.

Since the Uhlmann fidelity factorizes for product states,
\begin{align}
    F(\rho_\mathrm{A} \otimes \rho_\mathrm{B}, \rho'_\mathrm{A} \otimes \rho'_\mathrm{B})
    & = F(\rho_\mathrm{A}, \rho'_\mathrm{A}) F(\rho_\mathrm{B}, \rho'_\mathrm{B}),
    \label{eq:prod_relation}
\end{align}
the associated infinitesimal Bures metric is additive $ds_\mathrm{Bures}^2 (\mathrm{A} \otimes \mathrm{B}) = ds_\mathrm{Bures}^2(\mathrm{A}) + ds_\mathrm{Bures}^2(\mathrm{B})$.
Therefore, under the mean-field approximation, the total information metric density can be expressed as
\begin{align}
    \bar g^{(\mathrm{QFI})}_{ij}(T,\mu)
    & = \int_0^\Lambda \frac{d^3p}{(2\pi)^3} \,w\, \sum_{s=\pm} g^{(\mathrm{QFI}),s}_{ij} (\bm{p};T,\mu).
\label{eq:QFI_integrated_Bures}
\end{align}
where $w$ denotes the degeneracy factor associated with the quasiparticle spectrum.
In Eq.\,(\ref{eq:QFI_integrated_Bures}), $g_{ij}^{(\mathrm{QFI}),s}(p;T,\mu)$ is obtained by substituting $\bm{r} \to \bm{r}_\pm (p)$ in Eq.\,(\ref{eq:QFI_Bloch}), and thus the total metric is obtained by summing over the momentum and branch indices.
Equation~(\ref{eq:QFI_integrated_Bures}) defines the information metric used in the numerical calculations.
The degeneracy factor $w$ is in agreement with the mean-field thermodynamic potential, and its value is $w=4$ in this study.

\begin{figure}[t]
    \centering
    \includegraphics[width=1.0\linewidth]{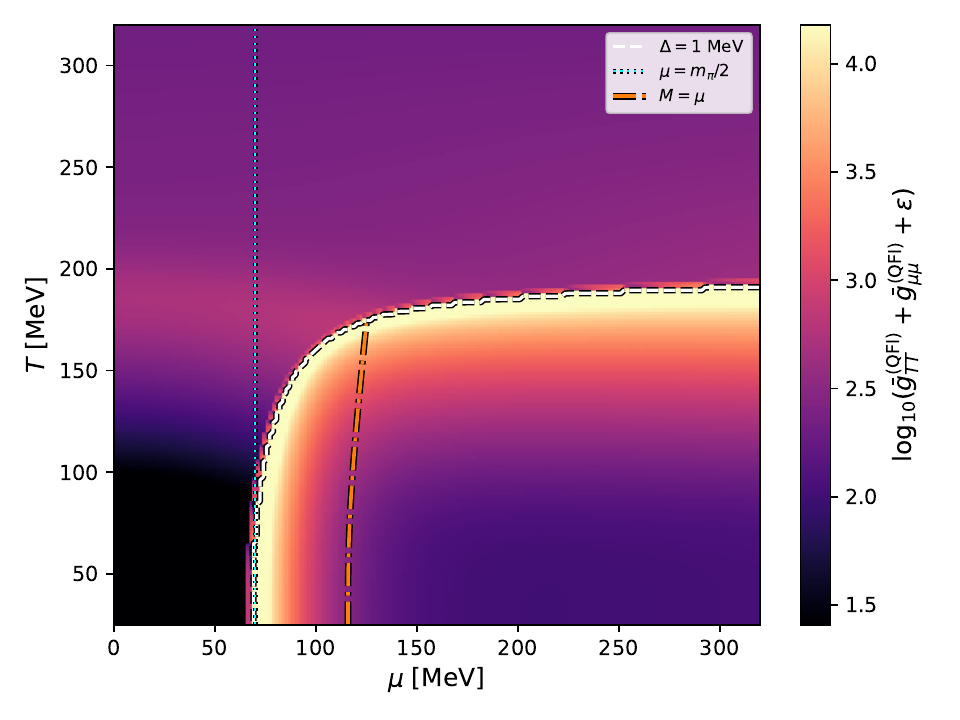}
    \caption{The QFI trace in the $T$--$\mu$ plane with the chiral crossover enhanced logarithmic representation, which is the same as shown in Fig.~\ref{fig:QFI}, but with a narrow color range.
    The color scale is adjusted to emphasize the weak enhancement around the chiral crossover.
    The dashed and dotted lines are the same as those shown in Fig.\,\ref{fig:M_Delta}.}
    \label{fig:QFI_etc}
\end{figure}
\begin{figure*}[t]
    \centering
    \includegraphics[width=0.49\linewidth]{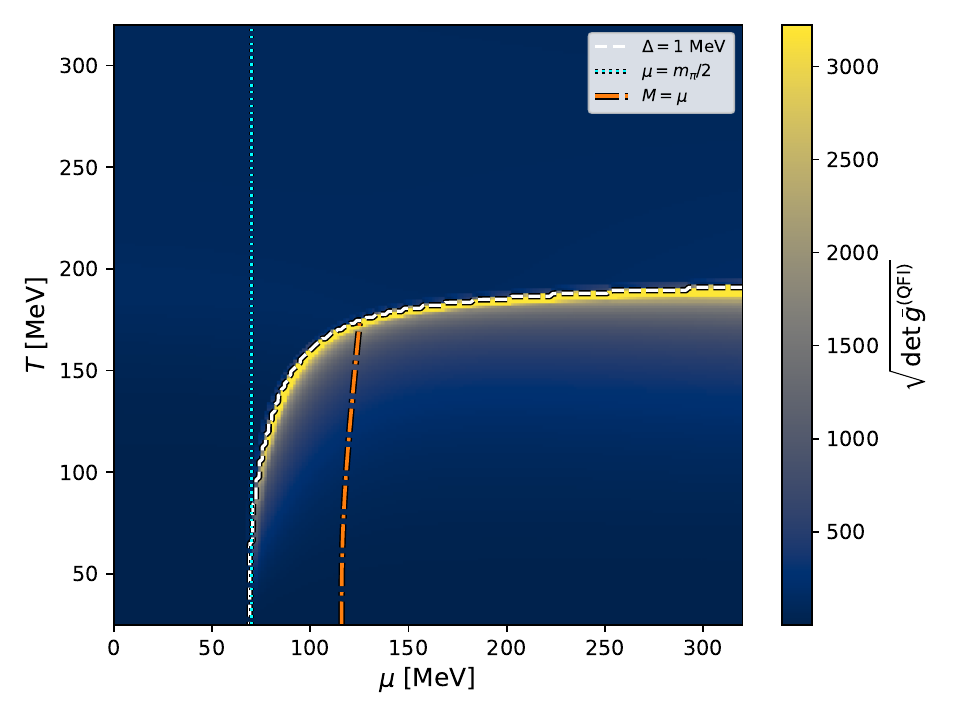}
    \includegraphics[width=0.49\linewidth]{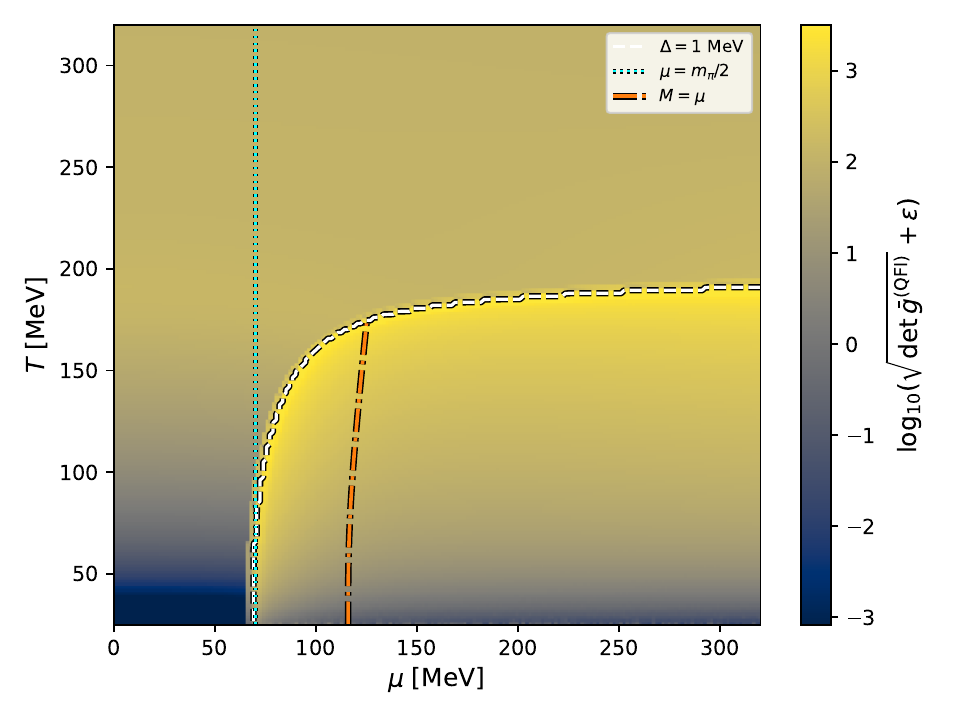}
    \caption{The metric-volume density based on the QFI geometry in the $T$--$\mu$ plane.
    The left and right panels show the linear and log scale results, respectively.
    The dashed and dotted lines are the same as those shown in Fig.\,\ref{fig:M_Delta}.}
    \label{fig:QFI_metric_volume}
\end{figure*}
\begin{figure}[t]
    \centering
    \includegraphics[width=1.0\linewidth]{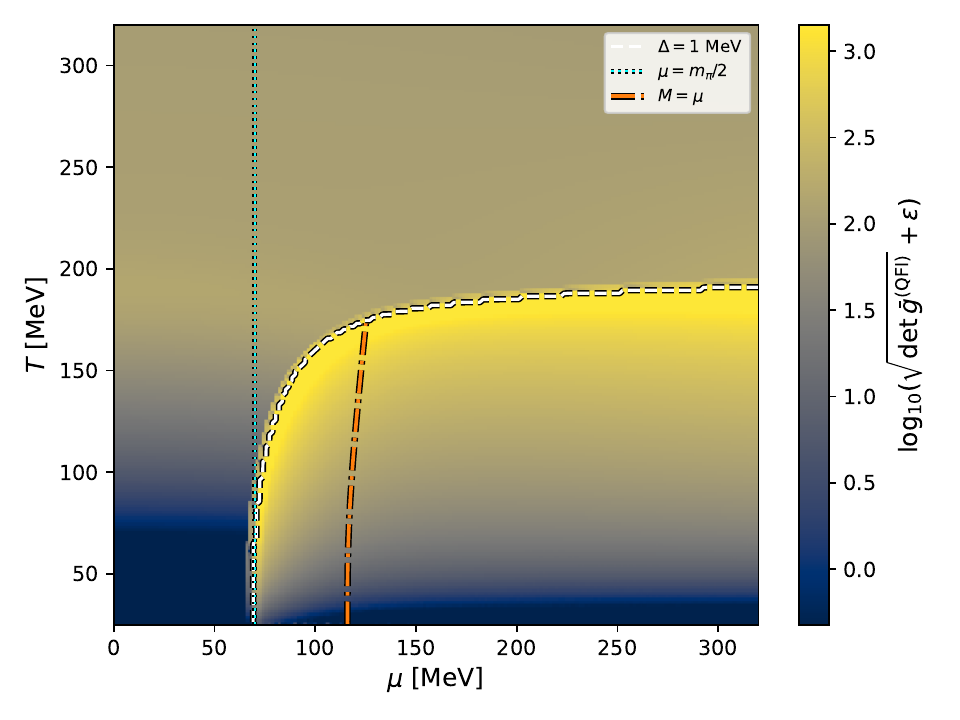}
    \caption{The metric-volume density in the $T$--$\mu$ plane with the chiral crossover enhanced logarithmic representation, which is the same as shown in Fig.~\ref{fig:QFI_metric_volume}, but with a narrow color range.
    The dashed and dotted lines are the same as those shown in Fig.\,\ref{fig:M_Delta}.}
    \label{fig:QFI_metric_volume_etc}
\end{figure}

\subsection{Metric-volume density}
In addition to the QFI trace, it is useful to consider a geometric quantity constructed from the determinant of the information metric.
For a two-dimensional thermodynamic manifold, the invariant volume element associated with the QFI metric is
\begin{align}
    dV &= \sqrt{\det \bar{g}^{(\mathrm{QFI})} } \, dT\, d\mu ,
\end{align}
where
\begin{align}
    \det \bar{g}^{(\mathrm{QFI})}
    &= \bar{g}^{(\mathrm{QFI})}_{TT} \bar{g}^{(\mathrm{QFI})}_{\mu\mu}
     - \left( \bar{g}^{(\mathrm{QFI})}_{T\mu} \right)^2.
\end{align}
Motivated by this geometric volume element, we define
\begin{align}
    V_g(T,\mu)
    &\equiv \sqrt{ \det \bar{g}^{(\mathrm{QFI})} (T,\mu)},
    \label{eq:metric_volume}
\end{align}
which is the standard Riemannian volume element associated with the QFI metric~\cite{amari2000methods,Braunstein:1994zz}. 
The quantity $V_g$ characterizes the local information volume density on the thermodynamic parameter manifold.
While the QFI trace measures the sum of the distinguishability responses along the coordinate directions, the metric-volume density measures the local area element.

\section{Numerical results}
\label{sec:numerical_results}
The QFI trace in the $T$--$\mu$ plane is shown in Fig.\,\ref{fig:QFI}.
There is a strong enhancement in the QFI trace along the color-superconducting phase boundary.
The enhanced region of the QFI trace follows the rapid variation of the diquark condensate.
Figure~\ref{fig:QFI_etc} shows the chiral-crossover-enhanced representation of the QFI trace in the $T$--$\mu$ plane, which is the same as shown in Fig.~\ref{fig:QFI}, but is displayed using a narrower color range.
In the figure, the weak enhancement associated with the chiral crossover becomes visible.

Figure\,\ref{fig:QFI_metric_volume} shows the metric-volume density constructed from the determinant of the QFI metric.
We observe a strongly enhanced region near the color-superconducting phase boundary.
In Fig.~\ref{fig:QFI_metric_volume_etc} which is the chiral-crossover-enhanced representation, a pronounced peak structure is again observed in the metric-volume density along the color-superconducting phase boundary.
By contrast, the enhancement around the chiral crossover is much weaker than that seen in the QFI trace, indicating that the metric-volume density is less sensitive to the chiral crossover.

Although the QFI trace and the metric-volume density capture the color-superconducting phase boundary and also the chiral crossover region, it is also interesting to examine its behavior in relation to the BEC--BCS crossover.
In dense two-color QCD, increasing $\mu$ induces a smooth crossover from a BEC of tightly bound diquarks to a BCS superfluid of weakly bound diquark Cooper pairs.
Since the BEC--BCS crossover is not associated with a thermodynamic singularity and occurs within the same color-superconducting phase, dramatic geometric signatures are not necessarily expected.

As shown in Figs.\,\ref{fig:QFI} and \ref{fig:QFI_metric_volume}, both the QFI trace and the metric-volume density exhibit relatively smooth behavior in the high-$\mu$ region beyond the color-superconducting phase boundary.  No pronounced
structure is observed near the expected BEC--BCS crossover.
A more detailed interpretation of these different geometric responses is given in the next section.

The present results suggest that the information geometric observables considered in this study are particularly sensitive to the rapid restructuring of the quasiparticle spectrum at the color-superconducting phase boundary.
In contrast, no sharp behavior associated with the BEC--BCS crossover, such as a peak structure, is observed.
Whether other geometric quantities can provide a clearer characterization of the BEC--BCS crossover remains an interesting question for future studies.

\section{Discussion}
\label{sec:discussions}
In this section, we discuss the origin of the different information-geometric responses observed near the color-superconducting phase boundary, the chiral crossover, and the BEC--BCS crossover.

To clarify the behavior of Eq.\,(\ref{eq:QFI_Bloch}), we substitute the explicit Bloch vector (\ref{eq:Bloch_vector}) into the QFI metric.  For each BdG branch, we then obtain
\begin{align}
    &g_{ij}^{(\mathrm{QFI}),\pm}
    \nonumber\\
    &=
    \sech^2\left(\frac{E_\Delta^\pm}{T}\right)
    \partial_i\left(\frac{E_\Delta^\pm}{T}\right)
    \partial_j\left(\frac{E_\Delta^\pm}{T}\right)
    \nonumber\\
    &+\tanh^2\left(\frac{E_\Delta^\pm}{T}\right)
      \frac{ \left( \xi_\pm \partial_i\Delta - \Delta\partial_i\xi_\pm \right)
             \left( \xi_\pm \partial_j\Delta - \Delta\partial_j\xi_\pm \right)
           }
           {\left(E_\Delta^\pm\right)^4},
\label{eq:QFI_decomposition}
\end{align}
where $x^i=(T,\mu)$, as in Eq.~(\ref{eq:ds_bures}).
The first term is controlled by the derivatives of $E_\Delta^\pm/T$.
Therefore, it becomes large when the quasiparticle energy changes rapidly.
The second term is controlled by the combination $\xi_\pm\partial_i\Delta-\Delta\partial_i\xi_\pm$.
This contribution becomes large when the balance between the pairing gap and the single-particle energy changes rapidly.

The derivatives of $\xi_\pm$ appearing in Eq.\,(\ref{eq:QFI_decomposition}) are given by
\begin{align}
    \partial_i\xi_\pm &= \frac{M}{E_p}\partial_i M \pm\delta_{i\mu}.
\end{align}
Near the color-superconducting phase boundary, $\Delta$ changes rapidly with $T$ and $\mu$.
The second term in the right-hand-side of Eq.\,(\ref{eq:QFI_decomposition}) directly contains $\partial_i\Delta$ through $\xi_\pm\partial_i\Delta-\Delta\partial_i\xi_\pm$.
Therefore, the QFI becomes quite sensitive to rapid changes in the diquark pairing state.
This observation provides a natural explanation for the
pronounced enhancement of the QFI trace near the
color-superconducting phase boundary.

The chiral crossover changes the constituent quark mass $M$ gradually.
Therefore, the derivatives entering Eq.~(\ref{eq:QFI_decomposition}) remain moderate compared with the rapid variation of the diquark gap near the color-superconducting boundary.

The absence of a pronounced structure near the BEC--BCS crossover can also be understood from Eq.~\eqref{eq:QFI_decomposition}.
The condition $\mu = M$ corresponds to a continuous rearrangement of the paired state within the same color-superconducting phase.
Unlike a phase boundary, it does not induce comparably rapid variations of either $E_\Delta^\pm$ or $\Delta$ itself.
Consequently, neither term in Eq.~\eqref{eq:QFI_decomposition} is expected to exhibit a pronounced enhancement near $\mu = M$.

The present results also provide a physical interpretation of the information-geometric quantities.
Equation~(\ref{eq:QFI_decomposition}) shows that the QFI is sensitive both to changes in the quasiparticle spectrum and to variations of the diquark gap.
Therefore, the enhancement of the geometric observables near the color-superconducting phase boundary can be understood as a consequence of the rapid change in the diquark pairing state.

\section{Summary}
\label{sec:summary}
In this study, we have investigated two-color QCD matter at finite temperature ($T$) and quark chemical potential ($\mu$) by using the two-flavor and two-color Nambu--Jona-Lasinio (NJL) model with the mean-field approximation from an information-geometric perspective.
Starting from the thermal states associated with the effective BdG Hamiltonians, we have constructed the Bures metric and the quantum Fisher information (QFI) on the thermodynamic parameter manifold spanned by $T$ and $\mu$.
To perform the numerical calculations, the quasiparticle spectrum was represented by a set of $2\times2$ Bogoliubov--de Gennes Hamiltonians reproducing the mean-field quasiparticle energies.

The numerical results indicate that the QFI trace exhibits clear peak structures near the color-superconducting phase boundary.
In addition, it shows a weaker enhancement in the chiral crossover region.
These results indicate that the QFI trace is particularly sensitive to rapid changes in the quasiparticle spectrum near the color-superconducting phase boundary.
We have also introduced the metric-volume density, which characterizes the local information volume element associated with the thermal-state manifold.
The metric-volume density also exhibits a pronounced peak near the color-superconducting phase boundary.
In contrast, neither the QFI trace nor the metric-volume density exhibits a sharp structure in the vicinity of the expected BEC--BCS crossover region, indicating that the BEC--BCS crossover produces a much weaker information-geometric response than the color-superconducting phase boundary.

Conventional susceptibilities require the selection of a specific order parameter beforehand, whereas the QFI trace and metric-volume density provide a unified measure of thermal-state distinguishability directly from the thermal state.
Although the susceptibilities and the QFI trace exhibit qualitatively similar enhancements around the color-superconducting phase boundary, their relationship is
not straightforward.
Susceptibilities characterize fluctuations of specific order parameters, whereas the QFI quantifies the distinguishability of neighboring thermal states.
Therefore, a simple one-to-one correspondence between these quantities is not expected a priori.
Clarifying how conventional susceptibilities are encoded in information-geometric observables remains an important open problem for future studies.

The present results suggest that such information-geometric observables can simultaneously capture the color-superconducting transition and a weaker signature of the chiral crossover.
Therefore, the present framework provides a new mixed-state geometric probe for dense QCD matter.
This feature may become particularly valuable in systems where the relevant order parameter is unknown, ambiguous, or difficult to compute.
The information-geometric quantities are constructed directly from the thermal state and therefore characterize the thermodynamic state itself.
In this sense, the present approach offers a state-based characterization of the phase structure.

The present formulation should be regarded as a mean-field quasiparticle information geometry rather than the full information geometry of two-color QCD matter.
The thermal state is represented in terms of reduced quasiparticle blocks, and the additivity of the information metric is employed within the mean-field
approximation.
Several extensions deserve further investigation.
A future comparison with a calculation including the full Nambu--Gor'kov internal structure would be useful for assessing the quantitative role of the additional degrees of freedom.
Incorporating collective fluctuations beyond the mean-field approximation is another important study, which may reveal how critical fluctuations modify information-geometric observables near the phase boundary.
Furthermore, because two-color QCD does not suffer from the sign problem at finite $\mu$, the present framework may open up the possibility of future comparisons with lattice QCD simulations.
Establishing such a connection may provide a bridge between mixed-state information geometry and  first-principle calculation of dense QCD matter.

\begin{acknowledgments}
This work is supported by the Japan Society for the Promotion of Science (JSPS) KAKENHI Grant Numbers JP22H05112.
\end{acknowledgments}

\appendix

\section{Susceptibilities}
\label{susceptibility}

\begin{figure*}[t]
    \centering
    \includegraphics[width=0.49\linewidth]{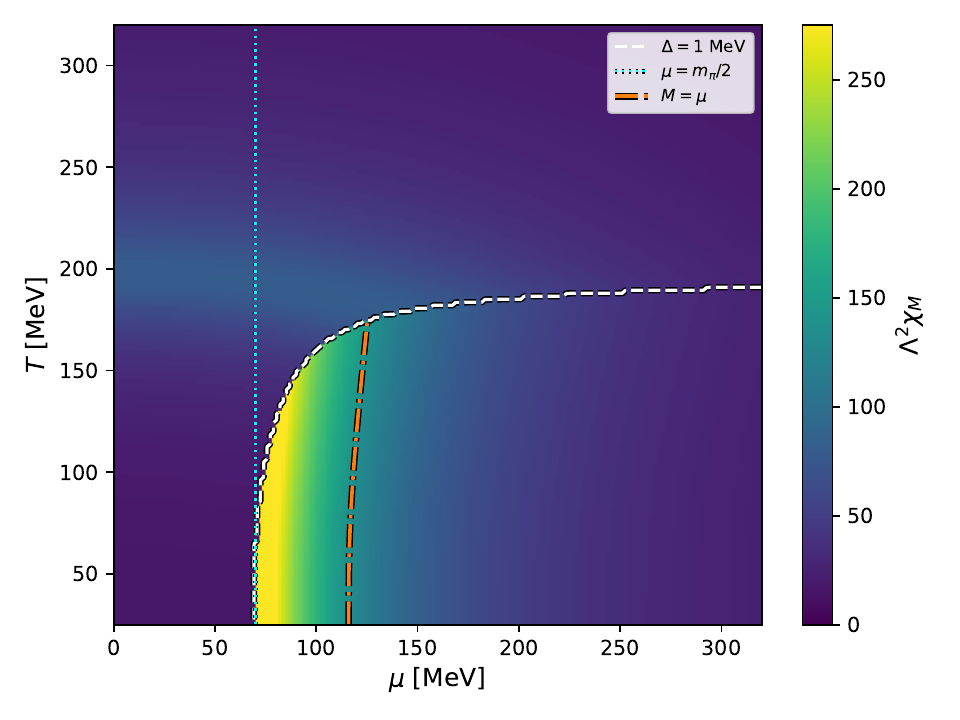}
    \includegraphics[width=0.49\linewidth]{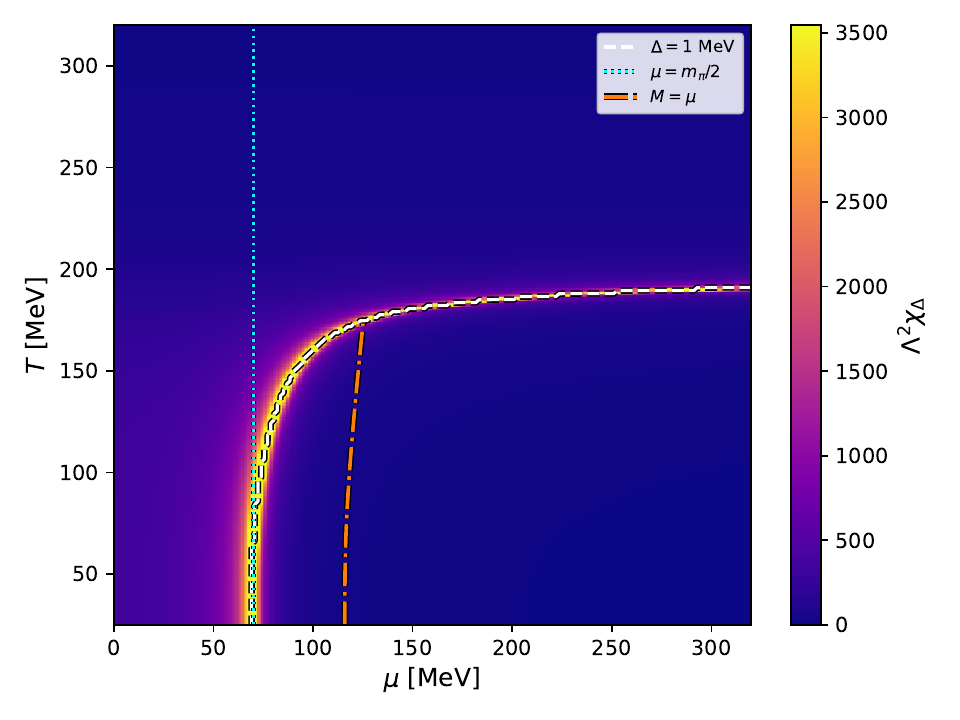}
    \caption{The chiral and the diquark susceptibility in the $T$--$\mu$ plane, where $\Lambda^2$ is multiplied to make the susceptibilities dimensionless.
    The dashed and dotted lines are the same as those shown in Fig.\,\ref{fig:M_Delta}.}
    \label{fig:susceptibility}
\end{figure*}

In this appendix, we evaluate the chiral and diquark susceptibilities from the curvature of the mean-field thermodynamic potential.

The Hessian matrix is defined as
\begin{align}
H &=
    \begin{pmatrix}
    \Omega_{MM}'' & \Omega_{M\Delta}'' \\
    \Omega_{M\Delta}'' & \Omega_{\Delta\Delta}''
    \end{pmatrix},
    \label{eq:Hessian}
\end{align}
with the curvature;
\begin{align}
    \Omega_{ab}''
    = \frac{\partial^2 \Omega_\mathrm{MF}}
           {\partial a \, \partial b},
\end{align}
where $a$ and $b$ take $M$ and $\Delta$, and $\Omega_\mathrm{MF}$ denotes the thermodynamic potential under the mean-field approximation.
Then, the susceptibility for the chiral condensate and the diquark condensate becomes
\begin{align}
    \chi_M
    &= (H^{-1})_{M M}
     = \frac{\Omega_{\Delta \Delta}''}
            {\Omega_{M M}'' \Omega_{\Delta\Delta}'' - \Omega_{M \Delta}''^{2}},
    \nonumber\\
    \chi_\Delta
    &= (H^{-1})_{\Delta\Delta}
     = \frac{\Omega_{M M}''}
            {\Omega_{M M}'' \Omega_{\Delta\Delta}'' - \Omega_{M \Delta}''^{2}}.
\label{eq:chi_dd}
\end{align}

Figure~\ref{fig:susceptibility} shows the chiral and the diquark susceptibility on the $T$--$\mu$ grid.
Although the diquark susceptibility exhibits a clear enhancement near the color-superconducting phase boundary, its peak remains finite.
This behavior is expected in numerical analysis because susceptibilities are evaluated on a finite $T$--$\mu$ grid, so the critical line is represented with finite resolution rather than as an exact singularity.

\section{Additivity of the QFI for the Reduced BdG Sectors}
\label{Justification}
Within the mean-field quasiparticle representation, the two BdG branches at fixed momentum are treated as independent sectors.
The corresponding effective state is therefore represented as
\begin{align}
    \rho(\theta) = \rho_+(\theta)\otimes\rho_-(\theta),
\end{align}
where each sector thermal state is separately normalized and $\theta=(T,\mu)$.

Since $\rho_+$ and $\rho_-$ are positive semidefinite, their spectral decompositions immediately imply
\begin{align}
    \sqrt{\rho_+ \otimes \rho_-}
    =
    \sqrt{\rho_+} \otimes \sqrt{\rho_-}.
\end{align}
For two product thermal states,
\begin{align}
    \rho &= \rho_+ \otimes \rho_-, ~~~
    \omega = \omega_+ \otimes \omega_-,
\end{align}
we obtain
\begin{align}
    \sqrt{\rho}\,\omega\,\sqrt{\rho}
    &= \left( \sqrt{\rho_+}\,\omega_+\,\sqrt{\rho_+} \right)
       \otimes
       \left( \sqrt{\rho_-}\,\omega_-\,\sqrt{\rho_-} \right).
\end{align}
Thus, we have
\begin{align}
    &\operatorname{Tr}
    \sqrt{\sqrt{\rho}\,\omega\,\sqrt{\rho}}
    \nonumber\\
    &=
    \operatorname{Tr}
    \sqrt{\sqrt{\rho_+}\,\omega_+\,\sqrt{\rho_+}}
    \times
    \operatorname{Tr}
    \sqrt{\sqrt{\rho_-}\,\omega_-\,\sqrt{\rho_-}}.
\end{align}
This shows that the root fidelity factorizes.
Consequently, the Uhlmann fidelity defined in Eq. (\ref{eq:Uhlmann_F}) is also multiplicative as
\begin{align}
    {\cal F}(\rho,\omega)
    &= {\cal F}(\rho_+,\omega_+) \, {\cal F}(\rho_-,\omega_-).
\end{align}

Using the multiplicativity of the Uhlmann fidelity and expanding for infinitesimally neighboring states, the Bures line element becomes additive over the independent BdG sectors as
\begin{align}
    ds_{\mathrm{B}}^2
    &= ds_{\mathrm{B},+}^2 + ds_{\mathrm{B},-}^2 .
\end{align}
The Bures metric is obtained from the infinitesimal expansion of the Uhlmann fidelity as in Eq.\,(\ref{eq:Bures_distance}), and thus the metric decomposes into independent sector contributions as
\begin{align}
    g^{(\mathrm{Bures})}_{ij} &= g^{(+)}_{ij} + g^{(-)}_{ij}.
\end{align}
Because the QFI metric is related to the Bures metric through Eq.\,(\ref{eq:QFI_metric_Bures}), the QFI metric is also additive over the two independent BdG sectors.
Therefore, within the mean-field approximation, we obtain 
\begin{align}
    g_{ij}^{(\mathrm{QFI})}
    &= g_{ij}^{(\mathrm{QFI}),+} + g_{ij}^{(\mathrm{QFI}),-}.
\end{align}
This establishes the branch additivity used in Eq.\,(\ref{eq:QFI_integrated_Bures}).

\section{Bloch representation of thermal state}
\label{Bloch_representation}
In this appendix, we briefly explain the Bloch representation of the thermal state.
Since the BdG Hamiltonian (\ref{eq:Hamiltonian_BdG}) represents a two-level system,
\begin{align}
    H_\pm(p) &= \bm{h}_\pm \cdot \bm{\sigma},
    \label{eq:BdG_H_s}
\end{align}
with
\begin{align}
    \bm h_\pm &= (\Delta,0,\xi_\pm),
\end{align}
we rewrite the thermal state to a more convenient form.

Using the Pauli-matrix algebra
\begin{align}
    \sigma_i\sigma_j &= \delta_{ij} I + i\epsilon_{ijk} \sigma_k,
    \label{eq:Pauli_matrix_algebra}
\end{align}
where $\epsilon_{abc}$ is the Levi-Civita symbol, we obtain the relation
\begin{align}
    (\bm{h}_\pm \cdot \bm{\sigma})^2 &= |\bm h_\pm|^2 I,
\end{align}
where $I$ is the identity matrix and
\begin{align}
    |\bm{h}_\pm| &= \sqrt{\xi_\pm^2 + \Delta^2}.
\end{align}
Consequently, we have
\begin{align}
    & (\bm{h}_\pm \cdot \bm{\sigma})^{2n}
      = |\bm{h}_\pm|^{2n}I,
    \\
    &(\bm{h}_\pm \cdot \bm{\sigma})^{2n+1}
      = (\bm{h}_\pm \cdot \bm{\sigma}) \, |\bm{h}_\pm|^{2n}.
\end{align}
With the relation, the numerator in Eq.\,(\ref{eq:density_matrix_pm}) can be evaluated by separating the even and odd powers in the series expansion as
\begin{align}
    e^{-H_\pm/T}
    &=
    \cosh \left(
        \frac{E_\Delta^\pm}{T}
    \right) I
    -
    \frac{\bm h_\pm\cdot\bm\sigma}
         {E_\Delta^\pm}
    \sinh \left(
        \frac{E_\Delta^\pm}{T}
    \right).
\end{align}
Taking the trace with $\operatorname{Tr}\,(\sigma_i)=0$,
the denominator becomes
\begin{align}
    \operatorname{Tr} \, e^{-H_\pm/T}
    &= 2 \cosh \left(
       \frac{E_\Delta^\pm}{T}
       \right),
\end{align}
and then we obtain
\begin{align}
    \rho_\pm
    &= \frac{1}{2}
    \left[
        I - \tanh \left(
            \frac{E_\Delta^\pm}{T}
            \right)
            \frac{\bm{h}_\pm \cdot \bm{\sigma}}
            {E_\Delta^\pm}
    \right].
    \label{eq:rho_Bloch}
\end{align}
Therefore, the Bloch representation of the thermal state becomes
\begin{align}
    \rho_\pm
    &= \frac{1}{2}
    \left(
        I + \bm{r}_\pm \cdot \bm{\sigma}
    \right),
    \label{eq:rho_Bloch2}
\end{align}
where the Bloch vector is identified as
\begin{align}
    \bm{r}_\pm
    &= - \tanh
        \left( \frac{E_\Delta^\pm}{T} \right)
        \left( \frac{\Delta}{E_\Delta^\pm},
         0,
         \frac{\xi_\pm}{E_\Delta^\pm}
         \right).
    \label{eq:Bloch_vector}
\end{align}

\begin{figure}[t]
    \centering
    \includegraphics[width=1.0\linewidth]{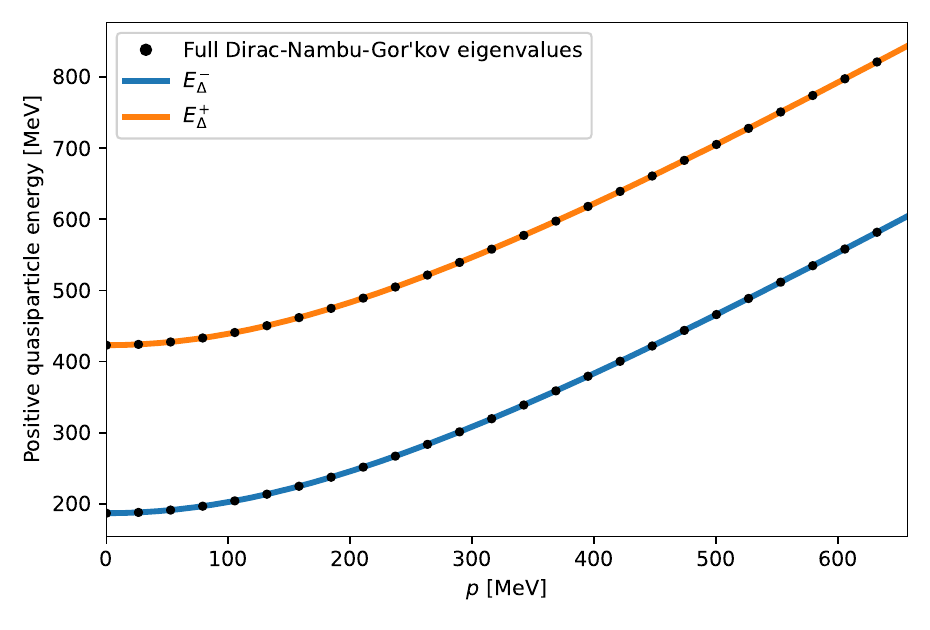}
    \caption{Positive quasiparticle eigenvalues obtained from the full Dirac-Nambu-Gor'kov Hamiltonian (circles) and the reduced $2 \times 2$ BdG blocks (solid curves), where we set $\mu=120$ MeV, $M=300$ MeV and $\Delta = 50$ MeV as an example.}
    \label{fig:full_energy}
\end{figure}

As a consistency check, we verified that the eigenvalue spectrum of the full mean-field Nambu-Gor'kov Hamiltonian coincides with the spectra of the two reduced BdG blocks as shown in Fig.\,\ref{fig:full_energy}.

\section{Quantum Fisher information metric}
\label{QFI_ap}
In this appendix, we briefly derive Eq.\,(\ref{eq:QFI_Bloch}), which is a standard result in quantum information theory.

For simplicity, we omit $\pm$ in the following explanation.
For later convenience, we first introduce the following;
\begin{align}
    r &= |\bm{r}|, ~~~ \bm{n}=\frac{\bm{r}}{r},
    \label{eq:def_1}
\end{align}
where $|\bm{n}| = 1$.
The thermal state (\ref{eq:rho_Bloch}) can be rewritten as
\begin{align}
    \rho &= \frac{1}{2} \left( I + r\,\bm{n} \cdot \bm{\sigma} \right).
    \label{eq:density_matrix_mod}
\end{align}
Using the Pauli-matrix algebra (\ref{eq:Pauli_matrix_algebra}), we obtain
\begin{align}
    (\bm{n} \cdot \bm{\sigma})^2
    & = n_a \, \sigma_a \, n_b \, \sigma_b
      = I.
\end{align}
Therefore, the eigenvalues of $\bm{n} \cdot \bm{\sigma}$ are $\pm 1$.
Thus, we obtain the eigenvalues of the thermal state as
\begin{align}
    \lambda_{\pm} &= \frac{1 \pm r}{2}.
\end{align}

To rewrite the thermal state to a convenient form, the following are useful.
Differentiating $\lambda_\pm$ using
\begin{align}
    \partial_i r
    &= \partial_i \sqrt{\bm{r} \cdot \bm{r}}
     = \frac{\bm{r} \cdot \partial_i \bm{r}}{r},
     \label{eq:dr}
\end{align}
we obtain
\begin{align}
    \partial_i\lambda_\pm &= \pm \frac{\bm{r}\cdot \partial_i \bm{r}}{2r}.
\end{align}
Note that we have
\begin{align}
    \langle \alpha | \partial_i \rho | \alpha \rangle &= \partial_i \lambda_\alpha,
\end{align}
in the spectral representation of the Bures metric (\ref{eq:Bures_metric}).
The spectral representation of the Bures metric contains both diagonal and off-diagonal contributions.
The diagonal terms depend only on the derivatives of the eigenvalues, whereas the off-diagonal terms arise from variations of the eigenvectors.
Therefore, both contributions are required to obtain the complete metric as follows.

The diagonal contribution is
\begin{align}
    g^{(\mathrm{Bures})}_{ij,\mathrm{diag}}
    &= \frac{1}{2} \sum_{\alpha = \pm}
       \frac{ \langle \alpha | \partial_i \rho | \alpha \rangle
              \langle \alpha | \partial_j \rho | \alpha \rangle }
            { 2 \lambda_\alpha },
\end{align}
and it becomes
\begin{align}
    g^{(\mathrm{Bures})}_{ij,\mathrm{diag}}
    &= \frac{1}{4} \frac{ (\bm{r} \cdot \partial_i \bm{r})
                          (\bm{r} \cdot \partial_j \bm{r}) }{ r^2 (1-r^2) }.
\end{align}
The off-diagonal contribution is
\begin{align}
    g^{(\mathrm{Bures})}_{ij,\mathrm{off}}
    &= \frac{1}{2} \sum_{\alpha \neq \beta}
       \frac{ \langle \alpha | \partial_i \rho| \beta \rangle
              \langle \beta| \partial_j\rho | \alpha \rangle }
            { \lambda_\alpha + \lambda_\beta }.
\end{align}

The derivative of Eq.\,(\ref{eq:density_matrix_mod}) becomes
\begin{align}
    \partial_i \rho = \frac{1}{2} (\partial_i \bm{r}) \cdot \bm{\sigma}
    &= \frac{1}{2}
       \Bigl[ (\partial_i r) (\bm{n} \cdot \bm{\sigma})
            + r (\partial_i \bm{n}) \cdot \bm{\sigma} \Bigr],
    \label{eq:drho}
\end{align}
where
\begin{align}
    \partial_i \bm{r} &= (\partial_i r) \, \bm{n} + r \, \partial_i \bm{n}.
    \label{eq:dvec_r}
\end{align}
Let $| + \rangle$ and $| - \rangle$ be eigenstates of $\bm{n} \cdot \bm{\sigma}$ as
\begin{align}
    (\bm{n} \cdot \bm{\sigma}) | \pm \rangle &= \pm | \pm \rangle,
\end{align}
$\bm{n} \cdot \bm{\sigma}$ is diagonal in the basis.
Therefore, we obtain
\begin{align}
    \langle + | \bm{n} \cdot \bm{\sigma} | - \rangle &= 0,
\end{align}
and the term proportional to $\partial_i r$ in Eq.\,(\ref{eq:dvec_r}) does not contribute to the off-diagonal matrix element.
Consequently, we obtain
\begin{align}
    \langle + | \partial_i \rho | - \rangle
    &= \frac{r}{2} \langle + | (\partial_i \bm{n}) \cdot \bm{\sigma} | - \rangle.
\end{align}
Since $\lambda_+ + \lambda_- = 1$, the off-diagonal contribution can be written as
\begin{align}
    g^{(\mathrm{Bures})}_{ij,\mathrm{off}}
    = \frac{1}{2} &\Big[ \langle + | \partial_i \rho | - \rangle
                         \langle - | \partial_j \rho | + \rangle
                       \nonumber\\
                       &~~~~~~~~~~~~
                       + \langle - | \partial_i \rho | + \rangle
                         \langle + | \partial_j \rho | - \rangle \Big].
\end{align}
Using the Hermiticity of $\partial_i\rho$ because $\rho$ is Hermitian, the second term is the complex conjugate of the first term.
Therefore, $g^{(\mathrm{Bures})}_{ij,\mathrm{off}} $ becomes
\begin{align}
    g^{(\mathrm{Bures})}_{ij,\mathrm{off}} 
    &= \Re \Big[ \langle + | \partial_i \rho | - \rangle
                 \langle - | \partial_j \rho | + \rangle \Big],
\end{align}
and it can be expressed as
\begin{align}
    &g^{(\mathrm{Bures})}_{ij,\mathrm{off}}
    \nonumber\\
    &= \frac{r^2}{4} \Re
       \Big[ \langle + | (\partial_i \bm{n}) \cdot \bm{\sigma} | - \rangle
             \langle - | (\partial_j \bm{n}) \cdot \bm{\sigma} | + \rangle \Big].
    \label{eq:g}
\end{align}
We inserted the completeness relation $| + \rangle \langle + | + | - \rangle \langle - | = I$ in Eq.\,(\ref{eq:g}), and we have
\begin{align}
    &\langle + | (\partial_i \bm{n}) \cdot \bm{\sigma} | - \rangle 
     \langle - | (\partial_j \bm{n}) \cdot \bm{\sigma} | + \rangle
    \nonumber\\
    & ~~~~~~~~~~~~
    = \langle + | (\partial_i\bm n) \cdot \bm{\sigma} \,
                (\partial_j \bm{n}) \cdot \bm{\sigma} | + \rangle
    \nonumber\\
    &~~~~~~~~~~~~~
    - \langle + | (\partial_i \bm{n}) \cdot \bm{\sigma} | + \rangle
      \langle + | (\partial_j \bm{n}) \cdot \bm{\sigma} | + \rangle.
      \label{eq:elements}
\end{align}
Using Pauli-matrix algebra,
\begin{align}
    (\bm{a} \cdot \bm{\sigma})(\bm{b} \cdot \bm{\sigma})
    &= (\bm{a} \cdot \bm{b})I + i(\bm{a} \times \bm{b}) \cdot \bm{\sigma},
\end{align}
we obtain
\begin{align}
    & ( \bm{n} \cdot \bm{\sigma})
      \left[(\partial_i \bm{n}) \cdot \bm{\sigma} \right]
    + \left[(\partial_i \bm{n}) \cdot \bm{\sigma} \right] (\bm{n} \cdot \bm{\sigma})
    \nonumber\\
    &= 2(\bm{n} \cdot \partial_i \bm{n})I.
    \label{eq:ndn}
\end{align}
Since $\bm{n} \cdot \bm{n} = 1$, we have
\begin{align}
    \partial_i(\bm n\cdot\bm n) = 2\bm n\cdot\partial_i\bm n = 0,
    \label{eq:dnn0}
\end{align}
and thus, Eq.\,(\ref{eq:ndn}) becomes
\begin{align}
    (\bm{n} \cdot \bm{\sigma}) \left[(\partial_i \bm{n}) \cdot \bm{\sigma} \right]
    + \left[(\partial_i\bm n) \cdot \bm{\sigma} \right] (\bm{n} \cdot \bm{\sigma})
    = 0.
\end{align}
This means that $(\partial_i \bm{n}) \cdot \bm{\sigma}$ anti-commutes with $\bm{n} \cdot \bm{\sigma}$.
Since $| + \rangle$ satisfies 
\begin{align}
    (\bm{n} \cdot \bm{\sigma}) | + \rangle = + | + \rangle,
\end{align}
and $\bm{n} \cdot \bm{\sigma}$ is Hermitian as
\begin{align}
    \langle + | (\bm{n} \cdot \bm{\sigma}) = \langle + |+,
\end{align}
we obtain
\begin{align}
    \langle + | (\partial_i \bm{n}) \cdot \bm{\sigma} | + \rangle
    &=\langle + | (\bm{n} \cdot \bm{\sigma}) \left[(\partial_i \bm{n}) \cdot \bm{\sigma} \right] | + \rangle
    \nonumber\\
    &= - \langle + | \left[(\partial_i \bm{n}) \cdot \bm{\sigma} \right] (\bm{n} \cdot \bm{\sigma}) | + \rangle
    \nonumber\\
    &= - \langle + | (\partial_i \bm{n}) \cdot \bm \sigma | + \rangle,
\end{align}
which is possible only if the matrix element vanishes.
Using it, the second term in Eq.\,(\ref{eq:elements}) vanishes.
Consequently, the off-diagonal contribution becomes
\begin{align}
    g^{(\mathrm{Bures})}_{ij,\mathrm{off}}
    &= \frac{1}{4} r^2 \partial_i \bm{n} \cdot \partial_j \bm{n}.
\end{align}
Using Eq.~(\ref{eq:dvec_r}) and (\ref{eq:dnn0}), we have
\begin{align}
    \partial_i\bm{r} \cdot \partial_j \bm{r}
    &= (\partial_i r)(\partial_j r) + r^2 \partial_i\bm{n} \cdot \partial_j \bm{n}.
\end{align}
Substituting Eq.~(\ref{eq:dr}) into this relation, we find the relation
\begin{align}
    r^2 \partial_i \bm{n} \cdot \partial_j \bm{n} 
    &= \partial_i \bm{r} \cdot \partial_j \bm{r} 
     - \frac{ (\bm{r} \cdot \partial_i \bm{r}) (\bm{r} \cdot \partial_j \bm{r}) }
            { r^2 }.
\end{align}
Hence, the off-diagonal contribution can be written as
\begin{align}
    g^{(\mathrm{Bures})}_{ij,\mathrm{off}} 
    &= \frac{1}{4} \left[ \partial_i \bm{r} \cdot \partial_j \bm{r}
     - \frac{ (\bm{r} \cdot \partial_i \bm{r}) (\bm{r} \cdot \partial_j \bm{r}) }
            { r^2 } \right].
\end{align}

Adding the diagonal and off-diagonal parts, we obtain
\begin{align}
    g^{(\mathrm{Bures})}_{ij} 
    &= \frac{1}{4} \left[ \partial_i\bm{r} \cdot \partial_j \bm{r}
     + \frac{ (\bm{r} \cdot \partial_i \bm{r}) (\bm{r} \cdot \partial_j \bm{r}) }
            { 1-r^2 } \right].
\end{align}
Using Eq.\,(\ref{eq:QFI_metric_Bures}), we finally obtain Eq.\,(\ref{eq:QFI_Bloch}).

\bibliography{ref.bib}
\end{document}